# Promising Properties of Sub-5 nm Monolayer MoSi$_2$N$_4$ Transistor

Junsheng Huang[1], Ping Li[1], Xiaoxiong Ren[1] and Zhi-Xin Guo[1,2,3*]

1. State Key Laboratory for Mechanical Behavior of Materials, Center for Spintronics and Quantum System, School of Materials Science and Engineering, Xi'an Jiaotong University, Xi'an, Shaanxi, 710049, China.
2. Key Laboratory of Polar Materials and Devices, Ministry of Education.
3. Department of Physics and Institute for Nanophysics, Xiangtan University, Xiangtan 411105, China

\* E-mail: zxguo08@xjtu.edu.cn

**Abstract**

Two-dimensional (2D) semiconductors have attracted tremendous interests as natural passivation and atomically thin channels that could facilitate continued transistor scaling. However, air-stable 2D semiconductors with high performance were quite elusive. Recently, extremely air-stable MoSi$_2$N$_4$ monolayer had been successfully fabricated [Hong *et al*., Science 369, 670 (2020)]. To further reveal its potential applications in the sub-5 nm MOSFETs, there is an urgent need to develop integrated circuits. Here we report first-principles quantum transport simulations on the performance limits of n- and p-type sub-5 nm monolayer (ML) MoSi$_2$N$_4$ metal-oxide-semiconductor FETs (MOSFETs). We find that the on-state current in the MoSi$_2$N$_4$ MOSFETs can be effectively manipulated by the length of gate and underlap (UL), as well as the doping concentration. Very strikingly, we also find that the n-type devices the optimized on-state current can reach up to 1390 and 1025 µA/µm for the high performance (HP) and low power (LP) applications, respectively, both of which satisfy the International Technology Roadmap for Semiconductors (ITRS) requirements. Whereas, the optimized on-state current can meet the LP application (348 µA/µm) for the p-type devices. Finally, we find that the $MoSi_2N_4$ MOSFETs have ultra-low subthreshold swing and power delay product, which have potential to realize the high speed and low power consumption devices. Our results show that $MoSi_2N_4$ is an ideal 2D channel material for future competitive ultrascaled devices.

## I. INTRODUCTION

The downscaling of field-effect transistors (FETs) to the sub-7 nm channel length is in great demand for integrated circuits in the next decade. [1,2] However, traditional silicon FETs are suffering from the challenges of the short-channel effect, increased leakage current, and unnecessary power consumption. [3–5] To better solve these problems, numerous researchers have focused their attention on two-dimensional (2D) materials. Compared with that of three-dimensional materials, the uniform thickness and smooth surfaces of 2D materials are beneficial for high-speed application due to their suppressed carrier scattering and possible trap generation. Moreover, the atomically thin thickness of 2D materials make them ideal gate electrostatics with diminished short-channel effects and leakage currents. [6–14]

To date, layered transition-metal dichalcogenides (TMDCs) such as $MoS_2$ are the mostly studied 2D semiconductor material in FETs. [15–19] For example, the bilayer $MoS_2$ transistor with a 1 nm carbon nanotube (CNT) gate had been experimentally fabricated and found to exhibit an excellent gate control ability with subthreshold swing (SS) of ~65 mV/dec and a remarkable on/off current ratio of $10^6$. [15] However, the on-state current (<250 μA μm$^{-1}$) of the 2D MoS$_2$ FETs is too low to meet ITRS [20] standard for both HP and LP applications, [15–19] owing to the small carrier mobility of MoS$_2$. On the other hand, some 2D materials such as 2D InSe and black phosphorene (BP) have much higher carrier mobility and excellent performance with high on-currents, which can satisfy the ITRS requirements for both HP and LP standards with the device scaled down to ~5 nm. [21–23] However, their instability in air is an issue, which greatly limits the device fabrications and applications. [21,23] Although numerous efforts have been devoted to explore new 2D materials, such as silicene, tellurene, Bi$_2$O$_2$Se, WSe$_2$, GeSe, ReS$_2$, BiN, AsP, [24–33] for the sub-5 nm FETs, the 2D material with both excellent device performance and air-environmentally stability is still scarce.

Recently, a new 2D material MoSi$_2$N$_4$ has been successfully synthesized by chemical vapor deposition (CVD). [34] This material is excellent ambient stability, e.g. it can be stable even in immersion in HCl aqueous solution for 24 hours. [34] MoSi$_2$N$_4$ is also a semiconductor, the band gap of which can be efficiently tuned by either strain or electric field. [35–37] In combination with the characteristics of highly controllable growth and good ohmic contact with electrodes, such as Ti, Sc, Ni and NbS$_2$, MoSi$_2$N$_4$ has great potential applications in the next-generation FET devices. [34,38,39]

Here we theoretically investigate the performance of n- and p-type sub-5 nm double-gated (DG) ML MoSi$_2$N$_4$ MOSFETs by using *ab initio* quantum transport calculations. We find that the transmission channel of ML MoSi$_2$N$_4$ locates in the MoN$_2$ layer sandwiched between two Si-N layers [as indicated in Fig. 1]. The main critical device properties, including the on-state current ($I_{on}$), SS, delay time (τ) and power-delay product (PDP), have been taken into consideration. More importantly, we also find that the on-state current of the optimized n-type ML MoSi$_2$N$_4$ MOSFET for HP is as high as 1390 μA/μm. The n-type ML MoSi$_2$N$_4$ MOSFET can meet the ITRS standards for HP and LP devices when scaled down to 3 nm and 1 nm, respectively. Whereas, the p-type ML MoSi$_2$N$_4$ MOSFET can only satisfy the ITRS LP requirement when it scaled down to 3 nm. Moreover, we find that the MoSi$_2$N$_4$ MOSFET has excellent gate controllability with a quite small SS, and its delay time and PDP are small enough to meet the ITRS HP and LP requirements.

## II. METHOD

The geometric optimization and electronic structures of monolayer MoSi$_2$N$_4$ were calculated by density-functional theory (DFT) with the projector augmented wave (PAW) method, which is implemented in the Vienna *ab initio* simulation package (VASP). [40–42] The exchange-correlation function was described based on the generalized gradient approximation (GGA) in the form of Perdew-Burke-Ernzerhof (PBE) parameterization. [43] The convergence standards of the atomic energy and positions are less-than 1×10$^{−6}$ eV per atom and 1×10$^{−2}$ eV Å$^{−1}$, respectively. The cutoff energy of the wave function was set to 500 eV. The Brillouin zone was sampled by 15×15×1 Monkhorst-Pack k-point mesh [44] for geometrical optimization and 21×21×1 for electronic states.

The transport properties were simulated based on the DFT method combined with the nonequilibrium Green's function (NEGF) formalism, using the Atomistix ToolKit (ATK) 2019 package. [45,46] The drain current at a given bias voltage $V_b$ and gate voltage $V_g$ was calculated through the Landauer-Büttiker formula: [47]

$$I(V_g, V_b) = \frac{2e}{h} \int_{-\infty}^{+\infty} \{T(E, V_b, V_g)[f_s(E - \mu_s) - f_d(E - \mu_d)]\} dE \quad (1)$$

where $T(E, V_b, V_g)$ is the transmission coefficient, $f_s$ and $f_d$ stands for the Fermi-Dirac distribution functions for the source and drain, respectively. $\mu_s$ and $\mu_d$ stands for the electrochemical potentials of the source and drain, respectively. In the calculations, the Tier 3 basis set was adopted with HGH pseudopotentials, and generalized gradient approximation (GGA) in the form of PBE function was utilized to represent the exchange and correlation interactions. Because of the heavily screened electron-electron interaction by doping carriers, DFT GGA-based single-electron approximation is good in the description of the device electronic structure. [22,48,49] The real-space mesh cutoff was chosen to be 75 hartree, and the k−point meshes [44] were set as Monkhorst-Pack 11×1×129 and 11×1×1 for the electrode region and the central region in the Brillouin zone. Moreover, the boundary condition along the transverse, vertical, and transport directions were set to be periodic, Neumann, and Dirichlet type, respectively. [50]

### III. RESULT AND DISCUSSION
#### A. Channel materials and device configuration

The atomic structure of ML MoSi$_2$N$_4$ is shown in Fig. 1(a), which can be regarded as a MoN$_2$ layer sandwiched between two Si-N layers [44]. The optimized lattice parameter of ML MoSi$_2$N$_4$ is 2.91 Å, in good agreement with previous results. [34] The band-decomposed charge-density distributions [Fig. 1(b)] on valence-band maximum (VBM) and conduction-band minimum (CBM) show that the CBM and VBM of monolayer $MoSi_2N_4$ are mainly contributed by the middle $MoN_2$ layer. Fig. 1(c) further shows that the valence and conduction bands of ML $MoSi_2N_4$ are mainly contributed by $d_z^2$ and $d_{x^2-y^2}$ orbitals of Mo atoms, and Si and N atoms contribute little to them [Fig. S1 within the Supplemental Material [51]]. This feature means that the middle $MoN_2$ layer would be responsible for the electron transmission, which is verified by the transmission eigenstates calculations as shown in Fig. 1(d). Since the transmission pathway in $MoN_2$ is protected by the two outside Si-N bilayers, the $MoSi_2N_4$ can be recognized as a natural micro-wire with conduction wire surrounded by insulator wire.

The performance of $MoSi_2N_4$ MOSFETs was further investigated in use of ATK package. In the simulation, double gates were adopted, and degenerately doped ML $MoSi_2N_4$ was selected as two-probe electrodes [Fig. 2(a)]. Compared with the single gate case, dual gates can obviously increase the competence of gate modulation. [52] The electron transport direction is along the zigzag direction of $MoSi_2N_4$, and the gate was assumed to be an ideal rectangle in the device model. In the calculation, the electrodes with length of about 1 nm was adopted and the dielectric constant of silicon dioxide was set to be 3.9. As indicated in Fig. 2(a), in the channel the gate region is usually shorter than the dielectric region, and the uncovered dielectric region is called underlap region. Previous studies found that the gate underlap, which is the spacer area between the gate and electrode, plays an essential role in the scalability of gate length for the modeled FETs. [22] The proper length of the gate underlap, $L_{UL}$, could improve the device performance. Thus one can define the channel length ($L_{ch}$) as the sum of the gate length ($L_g$) and twice the length of the underlap ($L_{UL}$), namely, $L_{ch} = L_g + 2L_{UL}$. In the calculation, the atomic compensation charges method was used for doping in the electrodes and the doping concentration of $1.0 \times 10^{13}$ cm$^{-2}$ was adopted, unless otherwise specified. According to the ITRS 2013 edition requirements for HP and LP standards of sub-5 nm device in 2028, we used 0.64 V as the supply voltage ($V_{dd}$) and 0.41 nm as the equivalent oxide thickness (EOT) of dielectric material (silicon dioxide). Various length of the gate underlap ranging from 0-4 nm was considered for a comprehensive investigation on the performance of MoSi$_2$N$_4$ MOSFETs.

### B. On-state current

On-state current ($I_{on}$) is a key parameter for evaluating the transition speed of a logic device. A high $I_{on}$ is advantageous for efficient applications such as high-performance servers with high switching velocity. One can calculate $I_{on}$ (HP) and $I_{on}$ (LP) by applying $V_g$(on/HP) = $V_g$(off/HP) ± $V_{dd}$ and $V_g$(on/LP) = $V_g$(off/LP) ± $V_{dd}$, respectively. In order to obtain the value of the on-state current the first step is making these DG ML MoSi$_2$N$_4$ MOSFETs reach the off-state current. Following the requirements of ITRS for the off-state current, the $I_{off}$ of HP and LP devices were set to 0.1 and $5 \times 10^{-5}$ µA/µm, respectively.

The calculated transfer characteristics of sub-5 nm gate-length MoSi$_2$N$_4$ MOSFETs are shown in Figs. 2(b)-2(g). As shown in Figs. 2(b)-2(g), all of the

MOSFETs have small enough source-drain leakage current to reach the off-state requirements for high-performance standards. This is own to the large band gap and simple energy band state in the conductive region. However, for the low-power application with $L_{UL}$ < 2 nm, the 1 nm gate-length device cannot reach the requirement of the off-state current due to the short channel effect. The introduction of the underlap region enlarges the effective channel length and improves the MOSFETs' ability in reaching the off-state current, especially for the device with gate length $L_g$ = 1 nm. Whereas, when the gate length is long enough, increasing UL has little effect on the device performance. This is because that the introduction of UL will inhibit the tunneling of carriers in the off state, especially for the devices with the short gate which had the relatively large tunneling effect. For the devices with long gate, the tunneling current was small, as well as the impact of UL. These phenomena can be noticeably observed by comparing Figs. 2(b) and 2(d), where the *I-V* curve for the 5 nm gate length device with different $L_{UL}$ are almost overlapped in Fig. 2(d). We additionally explored the effect of increasing doping concentration on the MOSFETs current, i.e., with $5.0\times10^{13}$ cm$^{-2}$ electron doping on the electrodes for the device of $L_{UL}$ = 4 nm. Compared with low doping-concentration case, the increase of doping concentration obviously increases the current under the same $V_g$.

We further summarized the values of $I_{on}$ for the sub-5 nm MOSFETs, as shown in Fig. 3. It is seen that $I_{on}$ generally monotonically increases with the gate length increasing. For the HP ML MoSi$_2$N$_4$ devices [Figs. 3(a) and 3(b)], the on-state current of high doping concentration case ($5.0\times10^{13}$ cm$^{-2}$) is much higher than that of low doping concentration ($1.0\times10^{13}$ cm$^{-2}$). It is seen that only the n-type MOSFET with high concentration can meet the ITRS HP requirement (900 µA/µm), reaching up to 1206 and 1390 µA/µm for the 3 and 5 nm gate-length, respectively. Nevertheless, the highest $I_{on}$ of the low doping concentration case reaches 817 µA/µm, fulfilling 91% of the ITRS standard [Fig. 3(a)]. As for the p-type MOSFETs with low concentration, the highest $I_{on}$ appears for a 3 nm gate-length MOSFET with $L_{UL}$ = 4nm [Fig. 3(b)]. Similar to the n-type MOSFETs, the $I_{on}$ increases with the doping concentration increasing, where the p-type 5 nm gate-length MOSFET could get an $I_{on} \approx 618$ µA/µm with doping concentration of $5.0\times10^{13}$ cm$^{-2}$. This value fulfills 69% of the ITRS standard. Note that the introduction of the underlap region can help to increase the $I_{on}$ for the HP ML MoSi$_2$N$_4$ devices in most cases.

As for the LP application, the $I_{on}$ of the n-type and p-type of the ML MoSi$_2$N$_4$ devices increases with the gate-length increasing from 1 to 3 nm, both of which can meet the ITRS HP standard (295 µA/µm) under certain $L_{UL}$ [Figs. 3(c) and 3(d)]. The largest $I_{on}$ of the n-type MOSFETs are 793 and 1025 µA/µm for the low doping concentration and the high doping concentration, respectively. The $I_{on}$ of n-type MOSFETs can satisfy the ITRS LP standard requirement even as the gate-length goes down to 1 nm with $L_{UL}$ = 4 nm (315 µA/µm) [Fig. 3(c)]. As for the p-type device, the minimum gate-length to fulfill the LP requirement is 3nm [Fig. 3(d)], where the largest $I_{on}$ (348 µA/µm) appears with 5 nm gate-length and $L_{UL}$ = 1 nm.

It is noticed that the increase of underlap does not always improve the on-state current of the device [as indicated in Figs. 3(c) and 3(d)]. This feature can be owing to two competing mechanisms: on one hand, the increase of underlap region makes the channel barrier longer, which decreases the transmission possibility and suppresses the short-channel effect (positive effect). On the other hand, the gate controlling capability of the underlap region becomes weaker with its length increasing, which would degrade the performance of the device (negative effect). These two conflicting effects imply that the length of the underlap should be optimized to obtain the highest on-state current.

To illustrate the function of underlap and modulation mechanism of gate more clearly, we have calculated the local density of state (LDOS) and transmission spectra of the 1 nm gate ML MoSi$_2$N$_4$ MOSFETs with different $L_{UL}$ [Fig. 4] for the HP case. Here we defined the maximum electron barrier height $\Phi_m$ as the energy barrier to transport from the source to the drain. As shown in Figs. 4(a)-4(c), the same off-state current of 0.1μA/μm, $\Phi_m$ is reduced from 0.26 eV at $L_{UL}$ = 0 nm to 0.19 and 0.17 eV at $L_{UL}$ = 2 and 4 nm, respectively. The calculated transmission spectra confirm the variation of transmission barrier [Fig. 4(d)]. When a voltage of 0.64 V is applied, the CBM of the ML MoSi$_2$N$_4$ in the channel region would move downward, leading to the on-state of the MOSFETs [Figs. 4(e)-4(g)]. From Fig. 3(a), $I_{on}$ with $L_{UL}$ = 4 nm is the highest, followed by the case with $L_{UL}$ = 2 nm, and $I_{on}$ in the one without UL is the lowest. This is because CBM becomes lower with the $L_{UL}$ increasing, which leads to a higher on-state current. Correspondingly, the on-state transmission spectra edge at $L_{UL}$ = 4 nm is the upmost within the bias window, which is consistent with the result of highest on-state current for $L_{UL}$ = 4 nm. [Fig. 4(h)]. The variation of the UL lead to different carrier barrier height $\Phi_m$ and barrier length, resulting in the difference of gate control.

### C. Gate control

The ability of gate control of the FET in the subthreshold region is usually described by subthreshold swing (SS), which impacts on the device performance and decides the operating voltage of the device. The definition of the SS is

$$\text{SS} = \frac{\partial V_g}{\partial (lgI_{DS})} \qquad (2)$$

where $I_{DS}$ is the drain current. Note that the smaller SS corresponds to the better gate control ability. Fig. 5 shows the calculated SS with different gate length ($L_g$) and underlap ($L_{UL}$). It is found that SS generally increases with the decreasing of the $L_g$ and $L_{UL}$. In addition, for the n-type device without underlap ($L_{UL}$=0), the SS value sharply increases from 57 to 166 mV/dec with $L_g$ downsizing from 5 to 1 nm. This feature shows that the underlap is beneficial to the reduce of SS. This is because that the devices with long $L_{UL}$ can reach the off-state current with smaller $V_g$, which lead to the larger slope in the $I$-$V_g$ figure. Such effect is more noticeable in the relatively short gate-length devices. For example, in the n-type case with $L_{UL}$ = 4 nm, the SS of $L_g$ = 1 nm MOSFETs can be reduced by 55% (from 166 to 75 mV/dec), while it can be only reduced by 23% (from 57 to 44 mV/dec) for $L_g$ = 5 nm. The similar phenomenon is also observed in the p-type case. It is noticed that the minimum SS of the n-type and p-type MoSi$_2$N$_4$ MOSFETs are 44 and 58 mV/dec, respectively. Both of them are smaller than the Boltzmann's tyranny (60 mV/dec) that is believed to be fundamental limit of SS in MOSFETs at room temperature. [22] This limit is suitable for classical transistors with long channel whose current is mainly composed of thermionic injection. In the MOSFETs with an ultrashort channel down to a few nanometers, the contribution of the tunneling current could help to make the value of SS fall below the Boltzmann's tyranny. Taking both the tunneling and thermionic currents into consideration, namely $I_{DS} = I_{tunnel} + I_{therm}$, the SS can be expressed as follows: [22]

$$SS = \frac{\partial V_g}{\partial (lgI_{DS})} = \left[\frac{r_{tunnel}}{SS_{tunnel}} + \frac{1-r_{tunnel}}{SS_{therm}}\right]^{-1} \quad (3)$$

where

$$r_{tunnel} = \frac{I_{tunnel}}{I_{DS}}, \quad SS_{tunnel} = \frac{\partial V_g}{\partial (lgI_{tunnel})}, \quad SS_{therm} = \frac{\partial V_g}{\partial (lgI_{therm})}$$

In the transistors with long channel, the $I_{tunnel}$ can be neglected, leading to $r_{tunnel} = 0$, SS = $SS_{therm}$, which has a fundamental limit of 60 mV/dec. In the transistors with ultrashort channel, the tunneling should be considered, the $r_{tunnel} \neq 0$, the SS can be below the Boltzmann's tyranny when the $SS_{tunnel}$ is small enough. The tunneling current is presented as $I_{tunnel} = e^{-w\sqrt{m^*\Phi_B}}$, where $\Phi_B$ is the average barrier height, and w is the width of the barrier. According to the LDOS maps, the $\Phi_B$ decrease rapidly with the change of $V_g$, leading to a great change of $I_{tunnel}$. Thus, the $SS_{tunnel}$ can be very small, which could help to make the value of SS below the limit of 60 mV/dec.

### D. Delay time and power consumption

We have additionally explored the property of switching speed in the MoSi$_2$N$_4$ MOSFETs, which is an essential figure of merit for a digital circuit. The switching speed can be characterized directly by the intrinsic delay time ($\tau$), as

$$\tau = \frac{C_g V_{dd}}{I_{on}} \quad (4)$$

where $C_g$ is total gate capacitance, defined as the sum of the channel capacitance ($C_{ch}$) and the gate fringing capacitance ($C_f$) per width. The $C_f$ is speculated to be two times of the intrinsic channel capacitance, and $C_{ch}$ can be calculated by the formula

$$C_{ch} = \frac{\partial Q_{ch}}{W \partial V_g} \quad (5)$$

with $Q_{ch}$ being the total charge in the central region and $W$ being the channel width, respectively. The calculated values of $C_g$ for the n- and p-type sub-5 nm ML MoSi$_2$N$_4$ MOSFETs in Tables SI and SII [within the Supplemental Material [51]], respectively. It is found that $C_g$ (0.099–0.14 for n-type, 0.102–0.149 fF/μm for p-type) of the MoSi$_2$N$_4$ MOSFETs is much smaller than either the HP (0.6 fF/μm) or LP (0.69 fF/μm) ITRS standard, respectively. Figs. 6(a) and 6(b) further show the values of $\tau$ for n- and p-type sub-5 nm MoSi$_2$N$_4$ devices as a function of $L_g$. It is shown that the intrinsic delay time $\tau$ of both the n- and p-type devices with various gate-length and $L_{UL}$ can fulfill the ITRS requirement (0.423 ps) for the HP devices. Also, all the intrinsic delay time $\tau$ can meet the ITRS LP standard (1.493 ps) for the LP devices, except for the one in n-type with $L_g$ = 1 nm and $L_{UL}$ = 2 nm due to the very low $I_{on}$ (44 μA/μm). Moreover, the value of $\tau$ under certain $L_g$ and $L_{UL}$ can be several (ten) times smaller than the HP (LP) ITRS standard, indicating the great potential applications in high switching speed of MoSi$_2$N$_4$ MOSFETs.

Another significant concern for FET applications is the switching energy cost by power delay product (PDP) which can be calculated by the formula

$$\text{PDP} = V_{dd} I_{on} \tau = C_g V_{dd}^2 \quad (6)$$

Figs 6(c) and 6(d) show the calculated PDP of n-type and p-type sub-5 nm MoSi$_2$N$_4$ devices as functions of $L_g$, respectively. In both cases, PDP monotonously decreases with increase of $L_{UL}$. In addition, PDP of both n-type (0.023–0.101 fJ/μm) and p-type (0.026–0.121 fJ/μm) sub-5 nm MOSFETs are much lower than the ITRS requirements for HP (0.24 fJ/μm) and LP (0.28 fJ/μm) standards. This feature verifies that the MoSi$_2$N$_4$ MOSFETs devices also have advantage of low-power consumption.

### E. Discussion

The comparison of the main parameters including on-state current, subthreshold swing, delay time and power-delay product of ML MoSi$_2$N$_4$ MOSFETs and other 2D MOSFETs with $L_g \lesssim 5$ nm based on *ab initio* quantum transport calculations for HP and LP devices is shown in Table I and II, respectively. There are some 2D MOSFETs with very high on-state current for HP application, such as Phosphorene (4500 μA/μm), BiH (2320 μA/μm), Tellurene (2114 μA/μm) and Arsenene (2030 μA/μm). However, they are all p-type MOSFETs and are not suitable for n-type doped, which hinders the application in Complementary Metal Oxide Semiconductor (CMOS). For the n-type MOSFETs, the performance of ML MoSi$_2$N$_4$ is better than other 2D materials, especially for the LP application.

Two-dimensional MoS$_2$ transistors have been extensively studied for many years, both in experimental and simulation ways. The electronic structures of MoS$_2$ and MoSi$_2$N$_4$ are very similar. However, the performance of ML MoSi$_2$N$_4$ transistor is much better than that of ML MoS$_2$ transistor. the current along the transport direction is defined as $I = Nev$, where N is the number of carriers and $v$ is the velocity of carriers and is defined as $v = \mu E$, where $\mu$ and $E$ are the carrier mobility and electric field, respectively. The on-state current is proportional to the carrier mobility. For 2D materials, the intrinsic carrier mobility can be calculated using the following equation:

$$\mu_{2D} = \frac{2e\hbar^3 C}{3k_B T |m^*|^2 E_1^2} \quad (7)$$

where $C$ is the elastic modulus, $m^*$ is the effective mass, $T$ is the temperature and $E_1$ is the deformation potential (DP) constant. The elastic modulus of ML MoSi$_2$N$_4$ is about 4 times high than that of MoS$_2$ (about 530 N/m for MoSi$_2$N$_4$ and 128 N/m for MoS$_2$, both for holes and electrons), which makes the higher carrier mobility of the monolayer MoSi$_2$N$_4$ (about 1227 cm$^2$ V$^{-1}$ s$^{-1}$ for holes and 288 cm$^2$ V$^{-1}$ s$^{-1}$ for electrons) than that of MoS$_2$ (about 200 cm$^2$ V$^{-1}$ s$^{-1}$ for holes and 152 cm$^2$ V$^{-1}$ s$^{-1}$ for electrons). [34] Therefore, the ML MoSi$_2$N$_4$ transistors have much higher on-state current than that of ML MoS$_2$.

TABLE I. Comparison of the upper performance limit of the ML MoSi$_2$N$_4$ MOSFETs with other 2D MOSFETs with $L_g \leqslant 5$ nm for HP devices.

| | Doping type | $I_{ON}(\mu A/\mu m)$ | SS(mV/dec) | $\tau$(ps) | PDP(fJ/$\mu$m) |
|---|---|---|---|---|---|
| MoS$_2$ [53] | n-type | 473 | 58 | 1.287 | 0.195 |
| | p-type | 440 | 46 | 0.396 | 0.096 |
| WSe$_2$ [54] | p-type | 1464 | 82 | 0.168 | 0.156 |
| Bi$_2$O$_2$Se [33] | n-type | 916 | 114 | 0.240 | 0.141 |
| | p-type | 585 | 96 | 0.375 | 0.140 |
| BiH [55] | p-type | 2320 | 77 | 0.020 | 0.029 |
| GeSe [52] | n-type | 518 | 130 | 0.124 | 0.041 |
| | p-type | 1703 | 60 | 0.054 | 0.059 |
| Phosphorene [22] | p-type | 4500 | 76 | 0.055 | 0.135 |
| Tellurene [29] | p-type | 2114 | 102 | 0.068 | 0.098 |
| Silicane [56] | n-type | 1374 | 65 | 0.042 | 0.037 |
| | p-type | 871 | 67 | 0.075 | 0.043 |
| Arsenene [57] | p-type | 2030 | 77 | 0.017 | 0.032 |
| MoSi$_2$N$_4$ | n-type | 1390 | 44 | 0.064 | 0.057 |
| | p-type | 618 | 64 | 0.140 | 0.055 |

TABLE II. Comparison of the upper performance limit of the ML MoSi$_2$N$_4$ MOSFETs with other 2D MOSFETs with $L_g \leqslant 5$ nm for LP devices.

| | Doping type | $I_{ON}(\mu A/\mu m)$ | SS(mV/dec) | $\tau$(ps) | PDP(fJ/$\mu$m) |
|---|---|---|---|---|---|
| MoS$_2$ [53] | n-type | 324 | 56 | 0.552 | 0.093 |
| | p-type | 425 | 46 | 0.411 | 0.075 |
| WSe$_2$ [54] | p-type | 1132 | 63 | 0.149 | 0.108 |
| ReS$_2$ [58] | p-type | 329 | 72 | 0.700 | 0.150 |
| BiH [55] | p-type | 179 | 67 | 0.168 | 0.018 |
| GeSe [52] | n-type | 274 | 90 | 0.320 | 0.055 |
| Phosphorene [22] | p-type | 857 | 85 | 0.193 | 0.108 |
| Tellurene [29] | p-type | 451 | 57 | 0.206 | 0.063 |
| Silicane [56] | n-type | 467 | 77 | 0.054 | 0.016 |
| | p-type | 378 | 67 | 0.136 | 0.012 |
| Arsenene [57] | p-type | 341 | 77 | 0.101 | 0.023 |
| MoSi$_2$N$_4$ | n-type | 1025 | 44 | 0.086 | 0.057 |
| | p-type | 355 | 70 | 0.265 | 0.060 |

## IV. CONCLUSION

To summarize, we have explored the performance limit of sub-5 nm n- and p-type ML MoSi$_2$N$_4$ MOSFETs by applying precise *ab initio* quantum transport simulations. We have found that the middle MoN$_2$ layer of MoSi$_2$N$_4$ is responsible for the electron transmission, and the variation of on-state current in the MoSi$_2$N$_4$ MOSFETs can be effectively manipulated by the length of gate and underlap, as well as the doping concentration. More importantly, a competing mechanism for the influence of underlap length on the on-state current has also been found, indicating there is an optimized on-state current for a certain assemble of gate length and underlap length. In addition, we have also found that the n-type devices the optimized on-state current can reach 1390 and 1025 µA/µm for the HP and LP applications, respectively, both of which satisfy the ITRS requirements. Whereas, the optimized on-state current can meet the LP application (348 µA/µm) for the p-type devices. Finally, we have found that the sub-5 nm $MoSi_2N_4$ MOSFETs can have unusually short intrinsic delay time and low power delay product compared with the standards of ITRS, which make the devices with high speed and low power consumption. Considering that $MoSi_2N_4$ is remarkably stable in air, its sub-5 nm MOSFETs devices with high-performance are expected to be widely realized in the near future.

## Acknowledgements

"J. S. H and P. L contributed equally to this work. We are grateful for useful discussions with Prof. Ying Guo. This work was supported by the National Key R&D Program of China (2018YFB0407600), National Natural Science Foundation of China (No. 12074301 and No. 12004295), Fundamental Research Funds for Central Universities (No. xzy012019062), and Open Research Fund of Key Laboratory of Polar Materials and Devices, Ministry of Education. P. L. also thanks China's Postdoctoral Science Foundation funded project (No.2020M673364).

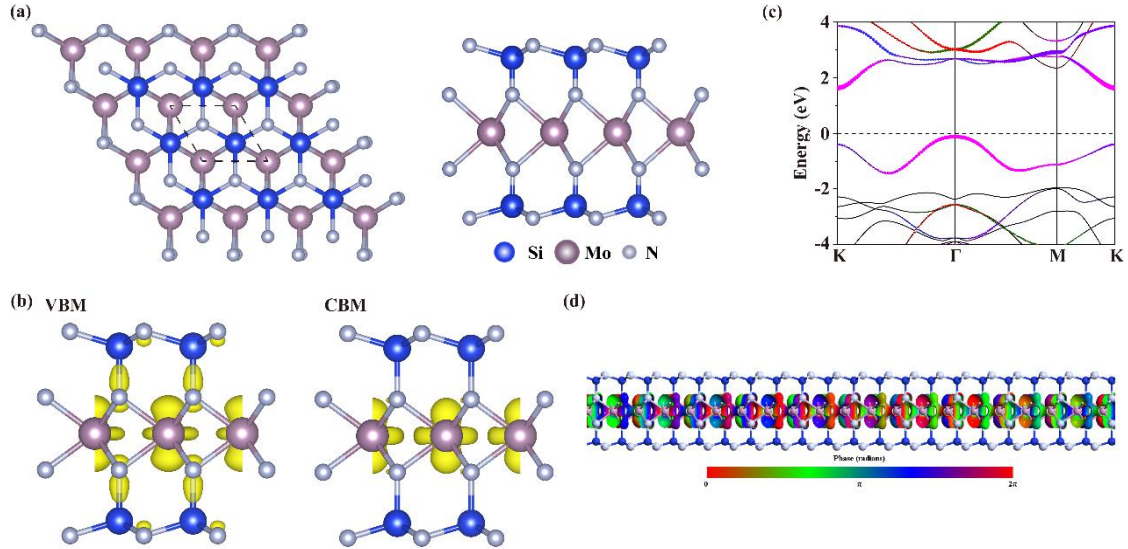

FIG. 1.  (a) Top view and side view of ML MoSi$_2$N$_4$, dashed lines represent the primitive unit cell of ML MoSi$_2$N$_4$. Blue, purple and white balls represent silicon, molybdenum and nitrogen atoms, respectively. (b) Band-decomposed charge-density distributions corresponding to valence-band maximum (VBM) and conduction-band minimum (CBM) of ML MoSi$_2$N$_4$. The isovalue is 0.02 e/bohr$^3$. (c) Electronic band structure of monolayer MoSi$_2$N$_4$. The different orbitals of Mo are mapped with different colors: Mo-d$_{xy}$ orbital, blue; Mo-d$_{yz}$ orbital, green; Mo –d$_{xz}$ orbital, red; Mo-d$_z{}^2$ orbital, magenta; Mo-d$_{x^2-y^2}$ orbital, purple and Mo-s orbital, orange. (d) Transmission eigenstates at K point (1/3,1/3) and E = 0.32 eV under onstate ($V_g$ = 0 V). The isovalue is 0.2 au.

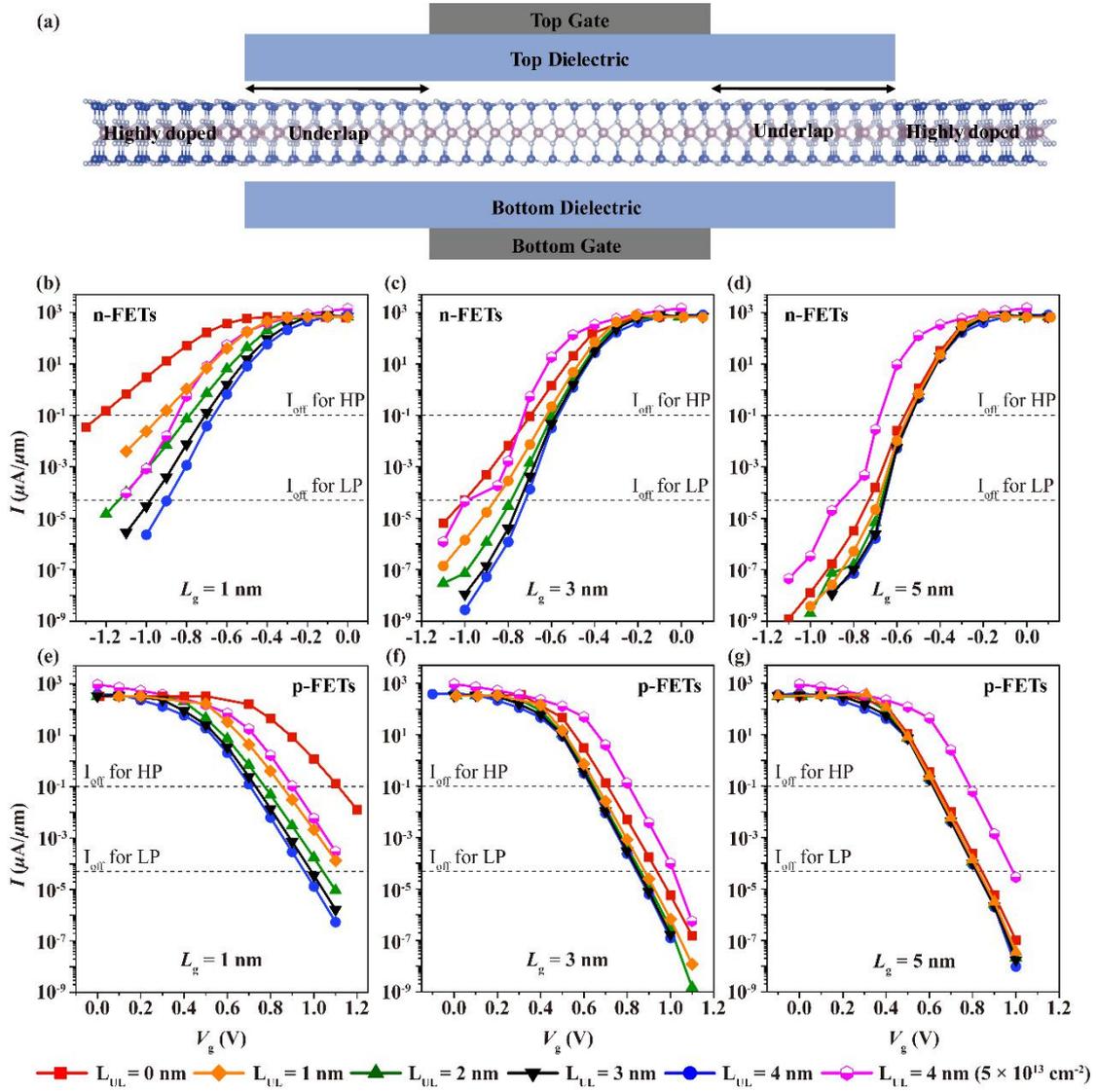

FIG. 2. (a) Schematic diagram of the DG ML MoSi$_2$N$_4$ MOSFETs. (b-g) $I$-$V_g$ characteristics of n- and p-type FETs with different gate lengths and $L_{UL}$ for $V_b$ = 0.64 V.

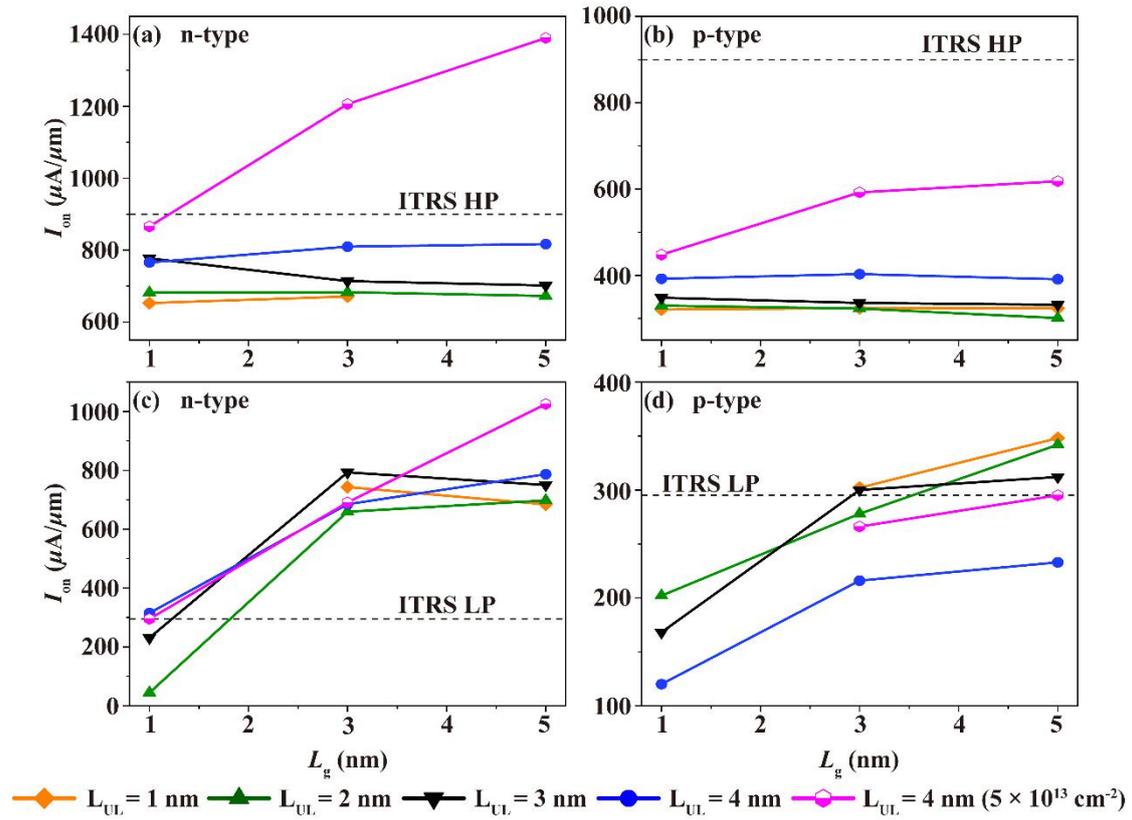

FIG. 3. On-state current as a function of the gate length for n- and p-type FETs with different $L_{UL}$. Black dashed lines represent the ITRS HP and LP requirements.

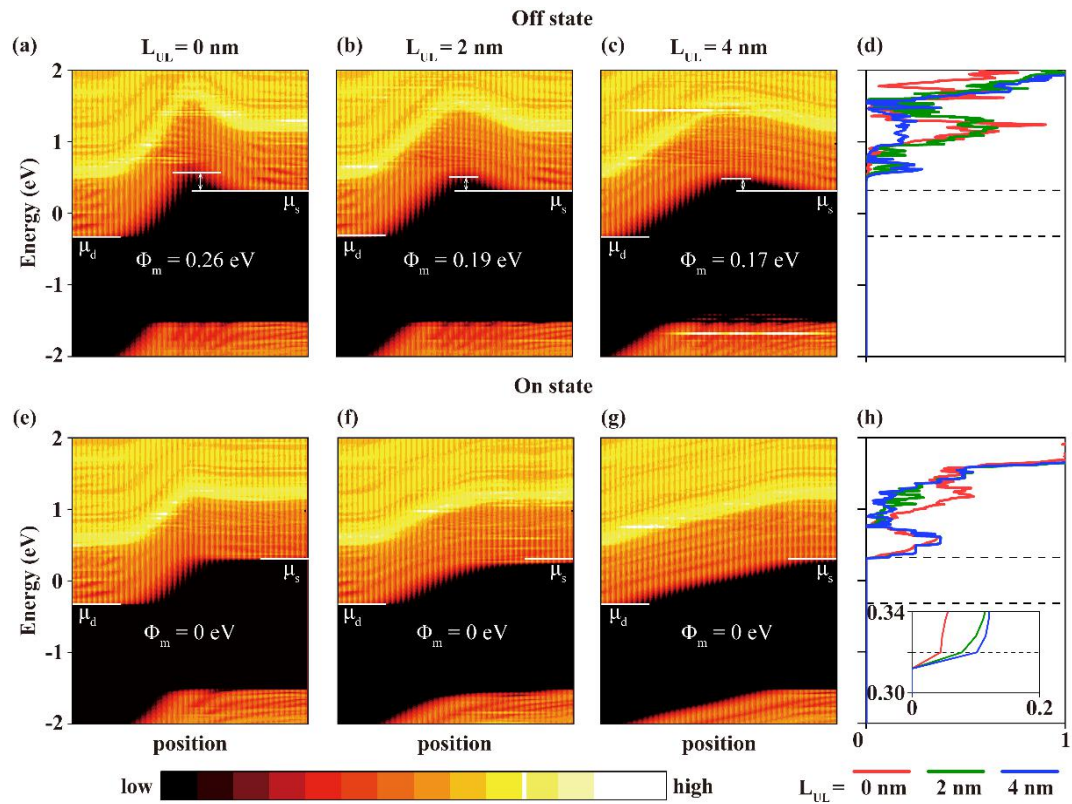

FIG. 4. Spatial resolved LDOS and transmission spectra for n-type FETs in the on and off states

with 1 nm gate length. Transmission spectra near the CBM of $MoSi_2N_4$ are shown in the inset in (h). $\mu_s$ and $\mu_d$ are the electrochemical potential of the source and drain, respectively.

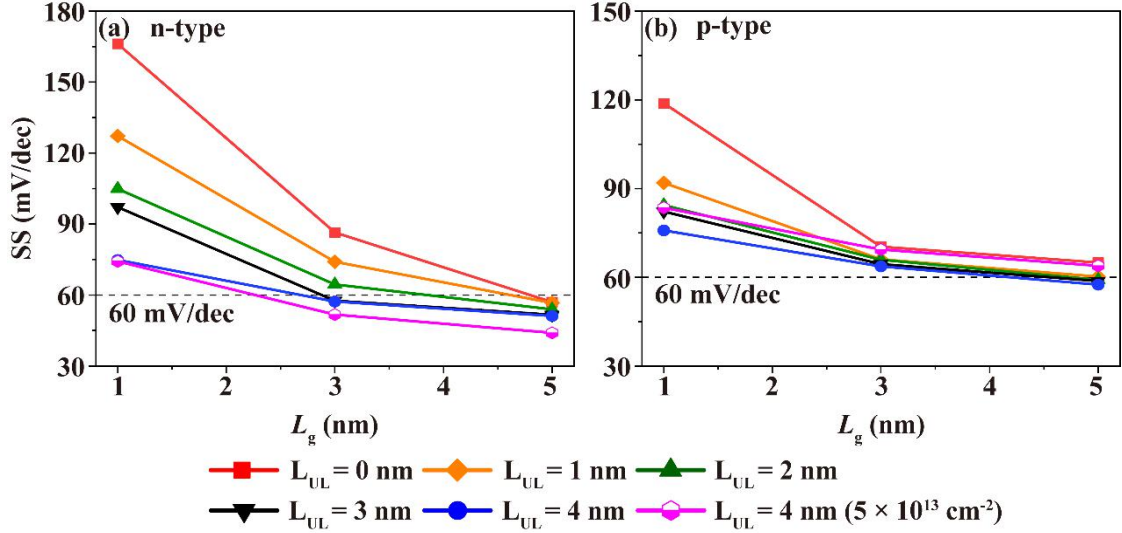

FIG. 5. Subthreshold swing (SS) as a function of gate length for n- and p-type FETs with different $L_{UL}$. Black dashed lines indicate the Boltzmann limit of 60 mV/dec for SS at room temperature.

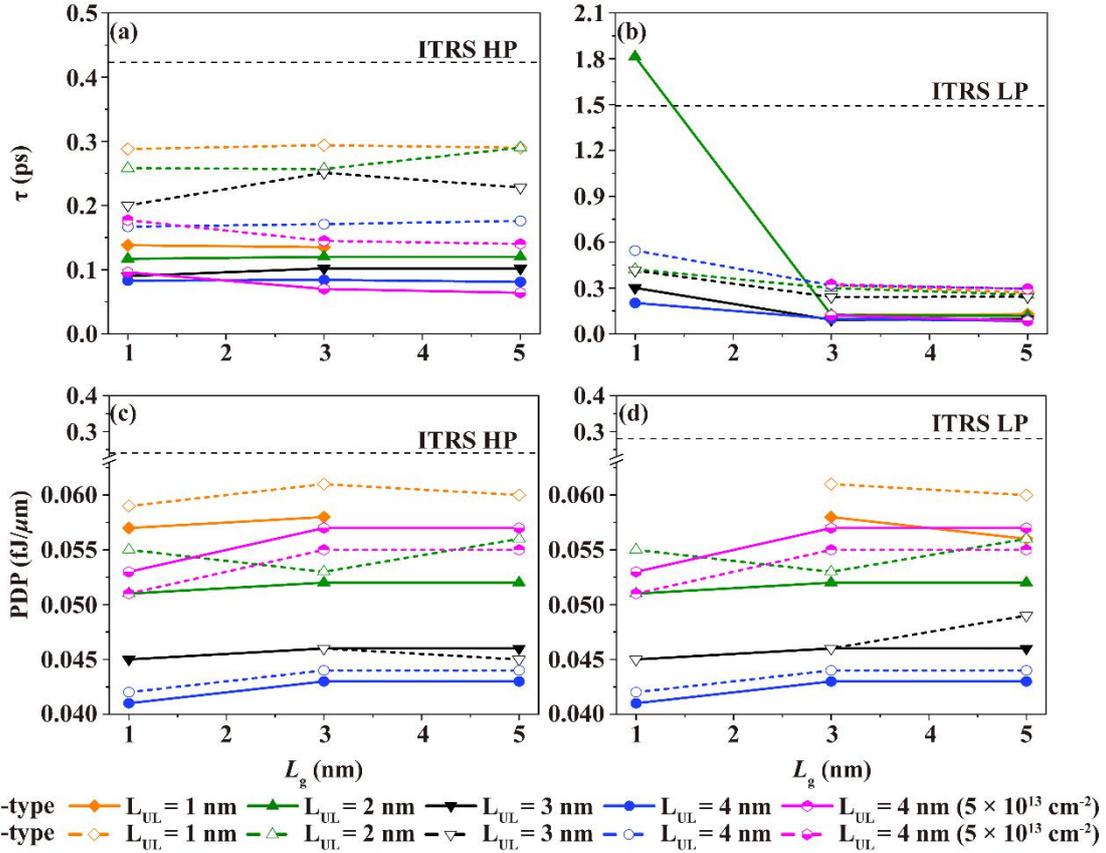

FIG. 6. Intrinsic delay time ($\tau$) (a–b) and power-delay product (PDP) (c–d) as a function of gate length for n- and p-type FETs with different $L_{UL}$. Black dashed lines are the ITRS HP and LP requirements for $\tau$ and PDP, respectively.